\documentclass[10pt,a4paper,twoside,dvips]{article}

\usepackage{vmargin,fancyheadings,multicol,ifthen,cite,
        epsfig,wrapfig,calc,dcolumn,doublespace}  % apalike
\usepackage{wgn2}

\begin{document}
\begin{WGNpaper}{twocol}{
\sectiontitle{Meteor science}
\title{Video observation of Geminids 2010 and Quadrantids 2011 by SVMN and CEMeNt.}
\author{Juraj T\'{o}th$^1$, Peter Vere\v{s}$^1$, Leonard Korno\v{s}$^1$, Roman Piffl $^2$, Jakub Koukal$^2$, \v{S}tefan
Gajdo\v{s}$^1$, Ivan Majchrovi\v{c}$^2$, Pavol Zigo$^1$, Martin
Zima$^2$, Jozef Vil\'{a}gi$^1$ and Du\v{s}an Kalman\v{c}ok$^1$\\
\thanks[Faculty of Mathematics, Physics and Informatics, Comenius University, Mlynska Dolina, 84248 Bratislava, Slovakia.{}\\
\thanks[CEMeNt{}\\
Email: {\tt toth@fmph.uniba.sk}}

\abstract{Since 2009 the double station meteor observation by the all-sky video
cameras of the Slovak Video Meteor Network (SVMN) brought hundreds of orbits. Thanks
to several amateur wide field video stations of the Central European Meteor Network
(CEMeNt) and despite not an ideal weather situation we were able to observe several
Geminid and Quadrantid multi-station meteors during its 2010 and 2011 maxima,
respectively. The presented meteor orbits derived by the UFOOrbit software account a
high precision of the orbital elements and are very similar to those of the SonotaCo
video meteor database.}

%\datereceived{Received 2011 XX XX}
}%
%
%==============================================================================
\section{Introduction}

The Geminids and Quandrantids belong to the most active and spectacular annual
meteor showers. Despite their high activity and relatively narrow orbital
distribution, no active comets have been associated with Geminids and Quadrantids as
the parent bodies yet. Geminids have very low perihelia ($\sim$\,0.15\,AU), moderate
geocentric velocities ($\sim34\,km\,s^{-1}$) and meteoroids seem to have high
density and strong internal consistence \cite{1}. Currently, asteroid (3200)
Phaethon is strongly favored as the parent body \cite{2,3}. Quadrantids exhibit a
high activity as well, with a sharp peak lasting only several hours. Their perihelia
lie at the Earth's orbit, inclinations are around $70\deg$. Among other parent body
candidates, asteroid (196\,256) 2003\,EH1 is having a most similar orbit to the
Quadrantids and is considered to be a dormant comet \cite{4}.

Both showers are well defined by many previous photographic,
radar, telescopic and visual surveys \cite{3,5,6}. Yet, the
photography has remained the most precise technique for the orbit
determination and atmospheric path definition. Recently, the video
observation with the high resolution digital cameras become
affordable and several meteor detection networks started operation
all over the world. Among them, the Slovak Video Meteor Network
(SVMN), operated by the Comenius University on the professional
level and Central European Meteor Network (CEMeNt) amateur network
consisting of several stations in Czech Republic and Slovak
Republic. Also we closely cooperate with the Polish Fireball
Network and Hungarian Meteor Network. Video observations are able
to detect fainter meteors than the classical photographic method,
obtain better time resolution of individual meteors and thanks to
available detection and analysis software, the data reduction is
fast.

\section{Slovak Video Meteor Network and CEMeNt}

The Slovak Video Meteor Network currently consists of two
semi-automated all-sky video cameras, developed and constructed at
the Astronomical and Geophysical Observatory in Modra of the
Comenius University (AGO). The first station is located at AGO,
the second one, remotely controlled, at the Arbor\'{e}tum
Mly\v{n}any (ARBO), in the distance of 80\,km. AGO is equipped
with the Cannon Fish-Eye lens (15\,mm, f/2.8), image intensifier
Mullard XX1332 and digital video camera DMK41AU02.AS
($1280\times960$\,px, angular resolution 8.5\,arcmin/pix, stellar
limiting magnitude +5.5, meteor limiting magnitude +3.5). The ARBO
station contains the same optical components but the analogue
camera Watec 902H2 ($720\times540$\,px, angular resolution
15\,arcmin/pix, stellar limiting magnitude +5, meteor limiting
magnitude +3). The third station is portable, it has the same
configuration as the station at AGO and was operating at the ARBO
site during the Quadrantids. The network web site is:
$http://www.daa.fmph.uniba.sk/meteor\_network.html$.

The CEMeNt network arouse out of amateurs observers initiation and
currently operates simultaneously with the SVMN. Observers among
the CEMeNt work independently ($http://cement.fireball.sk$). The
station at Dunajsk\'{a} Lu\v{z}n\'{a} is equipped with the Watec
902H2 Ultimate camera, XtendLan 2.8\,mm lens. It uses the AD
converter Canopus ADVC-55, the FOV is $114^\circ\times 85^\circ$.
Also the Stochov station has the same Watec camera and AD
converter as the Dunajsk\'{a} Lu\v{z}n\'{a} station but the
Fujinon lens (3.6\,mm). Its stellar limiting magnitude is +4.5,
for meteors approx. +1.5 and the FOV is $80^\circ\times60^\circ$.
Krom\v{e}\v{r}\'{i}\v{z} has the system consisting of the Watec
902 H2 camera ($720\times576$\,pix), Goyo GADN varifocal 3-8mm
lens, with the FOV $75^\circ\times60^\circ$, limiting stellar
magnitude +4.6 and meteor limiting magnitude +2.5. Vy\v{s}kov is
in fact the same station as Krom\v{e}\v{r}\'{i}\v{z} but in the
mobile form. The Marianka station consists of the Watec 902H2
Ultimate camera, Fujinon 2.9-8mm lens (F 0.95), the image is
obtained by the internal TV grabber of the Acer. The positions of
SVMN and CEMENT stations are depicted in Figure 1.

\begin {figure}
\centerline{\includegraphics[width=8.0cm]{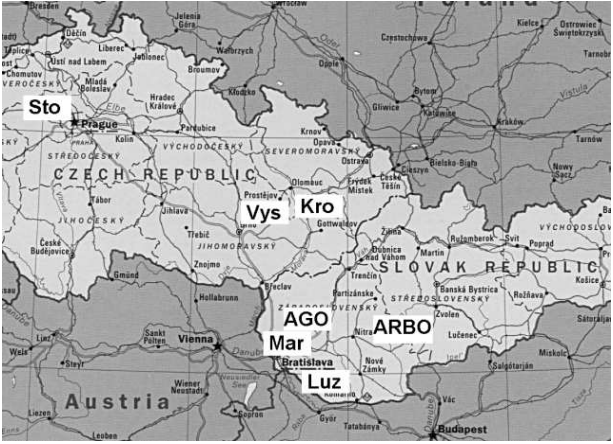}}
 \caption{Location of ground based video meteor
stations of SVMN (AGO, ARBO) and CEMeNt (Stochov, Vy\v{s}kov,
Krom\v{e}\v{r}\'{i}\v{z}, Marianka and Dunajsk\'{a}
Lu\v{z}n\'{a}).} \label{Figure 1}
\end{figure}

\section{Detection and data reduction}

The video signal is analyzed and detected by the UFOCapture software \cite{7} which
is able to recognize meteors and bolides. The meteor data had been processed by the
UFOAnalyzer and UFOOrbit software \cite{7}.

The meteor observations were performed during the maximum activity
of the Geminids (December 13-14, 2010) and Quadrantids (January
3-4, 2011), where 44, respectively 100 meteors were observed
simultaneously. In the data of the SVMN and CEMeNt we have found
35 Geminids and 66 Quadrantids.

There are several UFOOrbit parameters that can evaluate the
quality of the obtained meteor orbits. In order to separate hight
quality orbits, we set multiple constraints on the data set. Due
to the geometry of the incoming meteor trails we selected
individual meteor pairs in order to get the maximum precision of
the orbital elements. For the Geminids, we set the general quality
criteria for the orbits to Q2 -- internal condition of the
UFOOrbit \cite{9}. Finally, we obtained 10 Geminid meteor orbits.
Also, we selected Q3 quality criteria for Quadrantids and we
present 8 Quadrantid orbits.

\section{Results}

Orbits of Geminids (2010) and Quadrantids (2011) obtained during
the shower maxima are presented in Table 1 and Table 2. Also the
mean orbit is calculated as the arithmetic mean of each orbital
element with the corresponding standard deviation. In comparison
we used the SonotaCo data set of meteor showers observed above the
Japan during three years (2007-2009) and calculated the mean orbit
of the Geminids (121 orbits) and Quadrantids (39 orbits) as well.
Only SonotaCo orbits lying within the same range of the solar
longitude as our observed meteors have been taken into account (
$\lambda_{GEM\odot}\subset\langle261.49;261.79\rangle$ and
$\lambda_{QUA\odot}\subset\langle282.88;283.32\rangle$ ) The
comparison data were already filtered by the Q3 criterion in order
to process only the high quality orbit. As seen on the bottom of
the Table 1 and 2, our mean orbits are very similar of those of
the SonotaCo high quality subset according to its mean.

\begin{table*}[t]
\small
  \caption{Multi-station Geminids detected on December 13--14, 2010 by SVMN and CEMeNt video networks.
  Orbital elements, geocentric velocity and observing stations are presented. Stations: Dunajsk\'{a}
  Lu\v{z}n\'{a} (Luz), Vy\v{s}kov (Vys), Astronomical and Geophysical Observatory Modra (AGO).
  SonotaCo mean orbit from the solar longitude interval ($261.49^\circ - 261.79^\circ$) is also presented.}
  \vspace{0.2cm}
\begin{tabular}{c|c|c|c|c|c|c|c|c|c|c}
  \label{tab:tab1}
No  & a (AU) & q (AU) & e & i ($^\circ$) & $\omega$ ($^\circ$) & $\Omega$ ($^\circ$) & $\alpha$ ($^\circ$) & $\delta$ ($^\circ$) & $V_{g}$\,(km/s) & Station \\
\hline
1  &1.172  &0.152  &0.870  &20.86  &324.78  &261.49  &113.51  &32.09  &32.20  &AGO-Vys\\
2  &1.259  &0.159  &0.873  &21.42  &322.86  &261.52  &112.35  &32.62  &32.71  &AGO-Vys\\
3  &1.268  &0.146  &0.885  &23.29  &324.53  &261.57  &113.48  &32.59  &33.55  &AGO-Vys\\
4  &1.352  &0.145  &0.893  &23.16  &323.89  &261.58  &112.75  &32.43  &34.18  &AGO-Vys-Luz\\
5  &1.292  &0.144  &0.888  &22.62  &324.51  &261.58  &113.16  &32.23  &33.73  &AGO-Vys\\
6  &1.260  &0.147  &0.883  &22.77  &324.43  &261.59  &113.39  &32.49  &33.37  &AGO-Luz\\
7  &1.231  &0.150  &0.878  &21.13  &324.36  &261.61  &113.13  &32.01  &32.85  &AGO-Vys\\
8  &1.313  &0.135  &0.897  &22.85  &325.55  &261.62  &113.53  &31.79  &34.30  &AGO-Luz \\
9  &1.271  &0.155  &0.878  &19.42  &323.34  &261.71  &112.16  &31.50  &32.82  &Luz-Mar\\
10 &1.263  &0.145  &0.885  &20.86  &324.72  &261.79  &113.20  &31.57  &33.33  &Sto-Luz\\
\hline
mean &1.268 &  0.148  & 0.883 &  21.84  &324.30  &261.61  &113.07&  32.13  & 33.30 &  \\
st.\,dev  &0.048 &  0.007  & 0.008 &  ~1.28  &~~0.76  &~~0.09  &~~0.49&  ~0.41  & ~0.67 &   \\
\hline

SonotaCo & 1.279 & 0.149 & 0.884 & 22.69 & 324.03 & 261.69 & 113.24 & 32.45 & 33.47& 121 orbits\\
st.\,dev & 0.075 & 0.014 & 0.017 & ~2.49  & ~~1.45   & ~~0.08 & 0.76 & 0.78 & ~1.16 & \\
\end{tabular}
 \end{table*}

\begin{table*}
\small
  \caption{Multi-station Quadrantids detected on January 3--4, 2011 by SVMN and CEMeNt video
  networks. Orbital elements, geocentric velocity and observing stations are presented. Stations:
  Marianka (Mar), Dunaj\-sk\'{a} Lu\v{z}n\'{a} (Luz), Stochov (Sto), Krom\v{e}\v{r}\'{i}\v{z}
  (Kro), Astronomical and Geophysical Observatory Modra (AGO).
  SonotaCo mean orbit from the solar longitude interval ($282.88^\circ - 283.32^\circ$) is also presented.}\vspace{0.2cm}
  \label{tab:tab2}
  \begin{tabular}{c|c|c|c|c|c|c|c|c|c|c}
No  & a (AU) & q (AU) & e & i ($^\circ$) & $\omega$ ($^\circ$) & $\Omega$ ($^\circ$) & $\alpha$ ($^\circ$) & $\delta$ ($^\circ$) & $V_{g}$\,(km/s) & Station \\
  \hline
1   &2.040   &0.982   &0.518   &70.86  &175.56  &282.99  &226.65  &49.72   &39.40  &Mar-Sto-Kro \\
2   &2.608   &0.980   &0.624   &72.10  &172.51  &283.14  &228.67  &49.22   &40.87  &Kro-Sto \\
3   &2.433   &0.983   &0.596   &69.34  &179.15  &283.19  &227.39  &51.93   &39.32  &Kro-Sto \\
4   &2.779   &0.981   &0.647   &70.94  &173.73  &283.20  &229.36  &50.23   &40.51  &Kro-Sto \\
5   &2.477   &0.975   &0.607   &68.25  &167.54  &283.22  &233.18  &49.58   &38.90  &AGO-Kro \\
6   &2.645   &0.983   &0.628   &70.31  &177.64  &283.24  &227.85  &51.35   &40.06  &Ago-Sto-Luz \\
7   &2.471   &0.980   &0.604   &69.60  &171.83  &283.25  &230.46  &50.07   &39.52  &AGO-Sto \\
8   &2.921   &0.983   &0.663   &71.07  &178.44  &283.31  &227.45  &51.44  & 40.71  &Ago-Luz \\
\hline
mean  &2.547 & 0.981 & 0.611 & 70.31 & 174.55  &283.19 &  228.88 & 50.44 &  39.91 &  \\
st.\,dev &  0.264 & 0.003 & 0.044 & ~1.20 & ~~3.93  &~~0.10 &  ~~2.12 & ~1.00 &  ~0.73 &   \\
\hline
SonotaCo & 2.467 & 0.978 & 0.606 & 70.162 & 169.89 & 283.30 & 230.39 & 49.20 & 39.82 & 39 orbits \\
st.\,dev & 0.487 & 0.003 & 0.082 & ~2.59   & ~~2.98   & ~~0.16  & 2.36 & 0.93 & ~1.79 & \\
\end{tabular}
 \end{table*}

The ground projection of the individual meteor trails as seen by the multi-station
observation is depicted in Figure 2. The heliocentric orbits in the view
perpendicular to the ecliptic plane derived by the SonotaCo UFOOrbit software are
shown in Figure 3.

\begin {figure}
\centerline{\includegraphics[width=8.0cm]{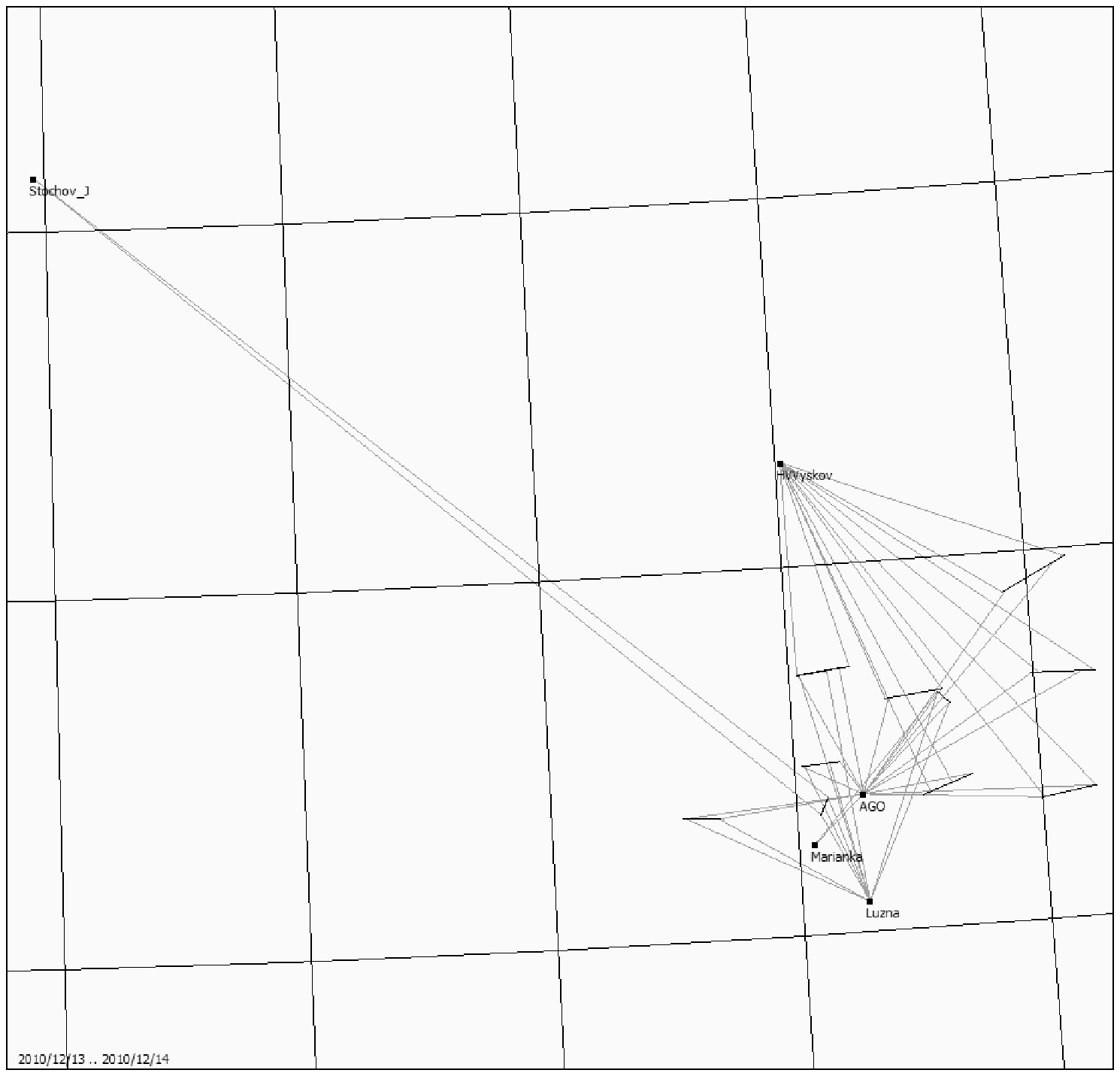}}
\centerline{\includegraphics[width=8.0cm]{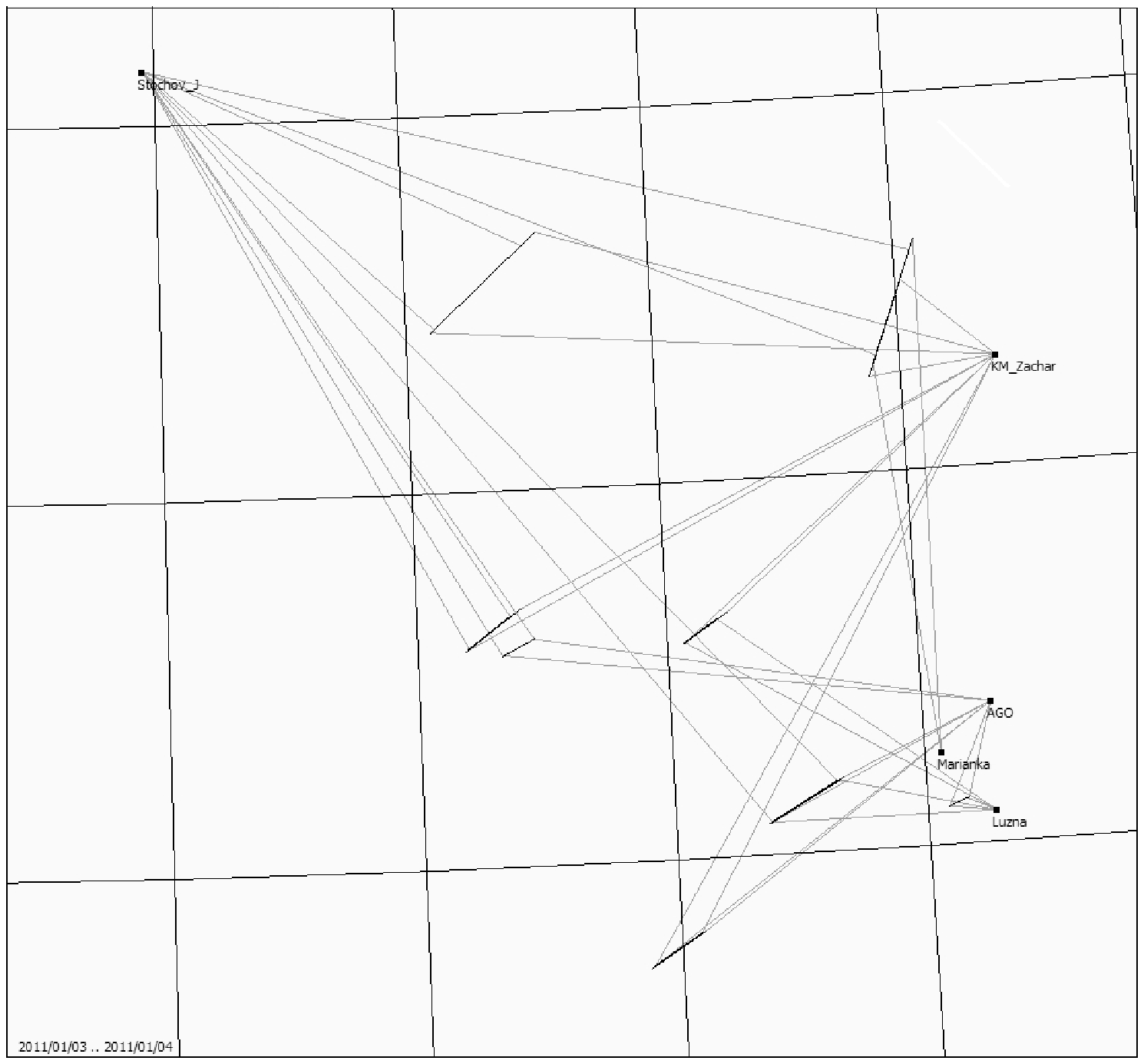}}

\caption{Ground projection of the meteor trails detected by SVMN
and CEMeNt. Upper image - Geminids 2010, lower image - Quadrantids
2011.} \label{Figure 1}
\end{figure}

\begin {figure}
\centerline{\includegraphics[width=7.0cm]{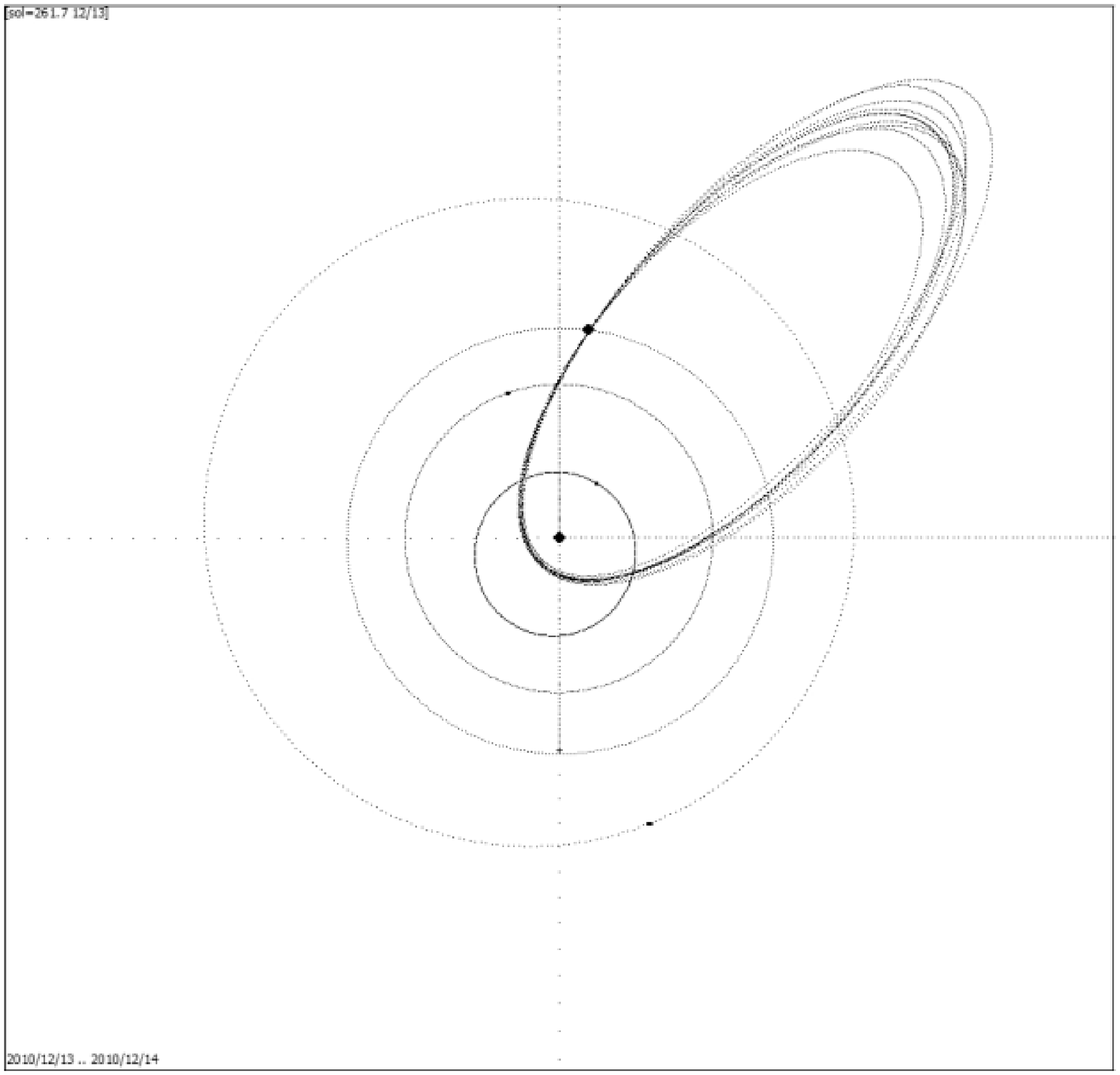}}
\centerline{\includegraphics[width=7.0cm]{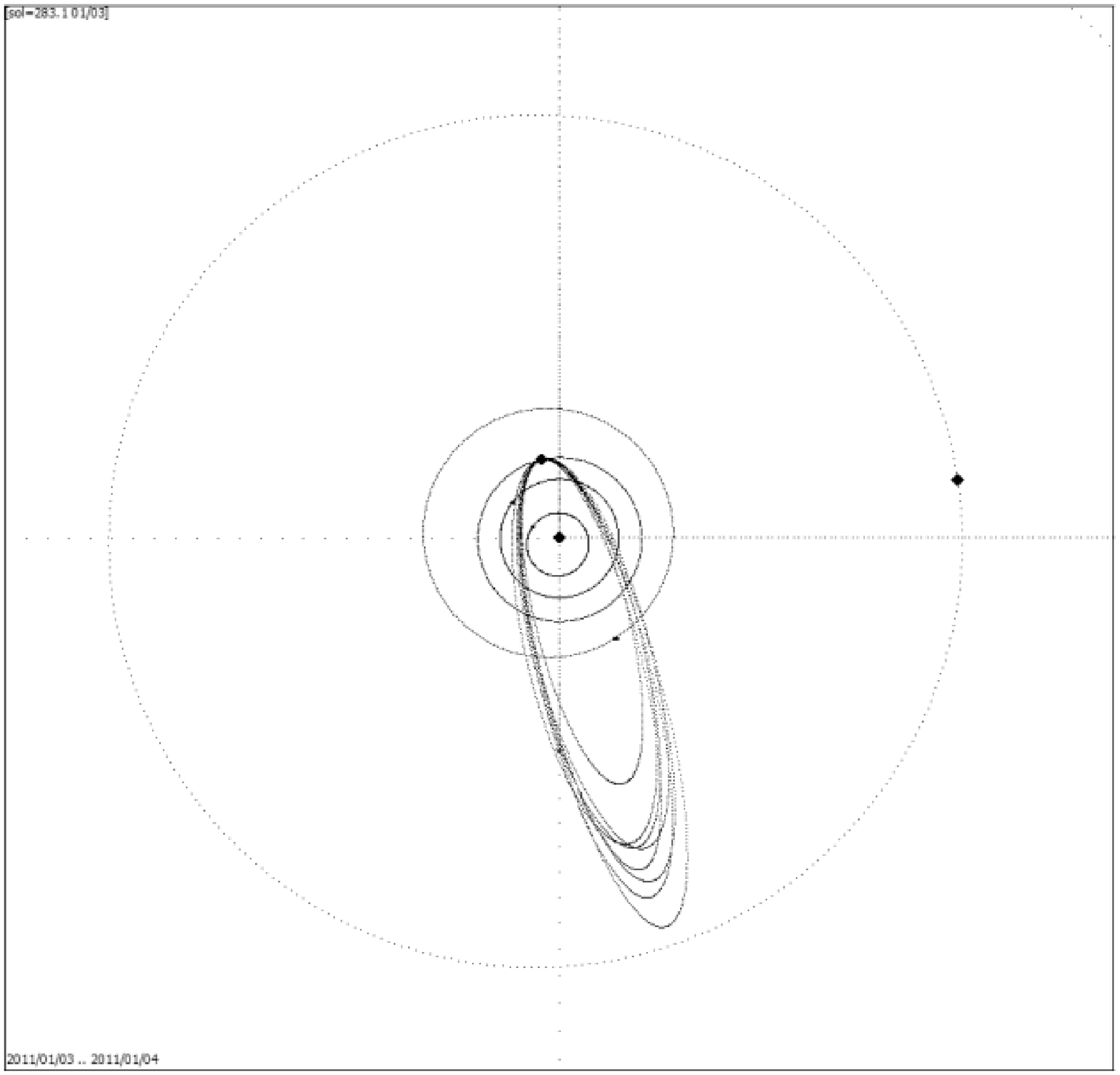}}

\caption{Orbits of multi-station meteors detected by video
stations, derived by UFOOrbit software. Upper image - Geminids
2010, lower image - Quadrantids 2011.} \label{Figure 2}
\end{figure}

To evaluate the quality of the derived meteor orbits, we employed
widely used Southworth-Hawkins D-criterion ($D_{SH}$) \cite{8}.
The crucial role in the criterion usage is the selection of the
reference nominal orbit. The obtained Geminid and Quadrantid
orbits were compared with respect to the mean orbits of the
showers derived from the SonotaCo video data (see Table\,1 and
Table\,2). Likewise, the parent body orbits were used for the
comparison. Individual ($D_{SH}$) values with respect to the
selected nominal orbits are shown in Table\,3 and Table\,4. Our
orbits are very similar to the SonotaCo mean orbits of the Geminid
and Quadrantid showers. Also the (3200) Phaethon orbit lies very
close to those of our derived orbits. On the other hand, the orbit
of the putative parent of the Quadrantids, (196\,256) 2003 EH1,
lies apparently somehow beyond of the mean orbit of the SonotaCo
and our data. The body undergone a series of close approaches to
Jupiter in past centuries, the last one in October 1972
($\sim0.28$\,AU). The 2003\,EH1 was much closer to the presented
Quadrantids about 170 years ago, when the orbits are integrated to
the past.

\begin{table}
\small
  \caption{Southworth-Hawkins D-criterion of Geminid orbits with respect to the SonotaCo
  mean orbit and putative parent body 3200 Phaethon.}\vspace{0.2cm}
  \label{tab:tab3}
 \begin{tabular}{l|c|c}
 No  & $D_{SH}$ (SonotaCo) & $D_{SH}$ (3200 Phaethon)\\
  \hline
1 &  0.036 &  0.043   \\
2 &  0.034 &  0.057   \\
3 &  0.012 &  0.037   \\
4 &  0.013 &  0.041   \\
5 &  0.008 &  0.031   \\
6 &  0.005 &  0.034   \\
7 &  0.028 &  0.039   \\
8 &  0.029 &  0.029   \\
9 &  0.059 &  0.065   \\
10&  0.034 &  0.035   \\
\hline
\end{tabular}
\end{table}

\begin{table}
\small
  \caption{Southworth-Hawkins D-criterion of Quadrantid orbits with respect to the SonotaCo
  mean orbit and putative parent body 2003 EH1 in years 2011 and 1840.}\vspace{0.2cm}
  \label{tab:tab4}
\begin{tabular}{l|c|c|c}
 No & SonotaCo  & 2003\,EH1 (2011)& 2003\,EH1 (1840) \\
  \hline
1   &   0.10 & 0.24 & 0.18 \\
2   &   0.05 & 0.21 & 0.05 \\
3   &   0.10 & 0.23 & 0.15 \\
4   &   0.06 & 0.21 & 0.04 \\
5   &   0.04 & 0.23 & 0.13 \\
6   &   0.09 & 0.22 & 0.11 \\
7   &   0.02 & 0.21 & 0.11 \\
8   &   0.11 & 0.23 & 0.07 \\
\hline
\end{tabular}
\end{table}

%\subsection{Beginning and terminal heights}

The beginning and the terminal heights as a function of the
absolute brightness of Geminids and Quadrantids are presented in
Figure \ref{Figure 4} and in the equations (\ref{eq:hei1}) and
(\ref{eq:hei2})

\begin{equation}
\begin{array}{l}
\displaystyle H_{B}=96.4(\pm 1.3) + 1.2(\pm 1.8)~M_{A} \\
\displaystyle H_{E}=84.2(\pm 0.9) + 2.8(\pm 1.3)~M_{A},
\end{array}
\label{eq:hei1}
\end{equation}

\begin{equation}
\begin{array}{l}
\displaystyle H_{B}=95.6(\pm 2.4) + 0.2(\pm 0.9)~M_{A} \\
\displaystyle H_{E}=81.1(\pm 5.9) + 1.1(\pm 2.4)~M_{A},
\end{array}
\label{eq:hei2}
\end{equation}

\noindent where $H_{B}$ stands for the beginning height, $H_{E}$ for the terminal
height and $M_{A}$ for the absolute brightness. The brightest Geminid meteor (Figure
\ref{Figure 4}) was not used in linear fit (\ref{eq:hei1}). It seems to be a special
case of solid meteoroid. However, the beginning heights do not change too much,
which is consistent with the results obtained by \cite{10}. Similarly, Quadrantids
also show stable beginning heights vs. brightness, at least in the observed
interval. Naturally, terminal heights decrease with the increasing brightness in
both meteor showers.

\begin {figure}
\centerline{\includegraphics[width=8.5cm]{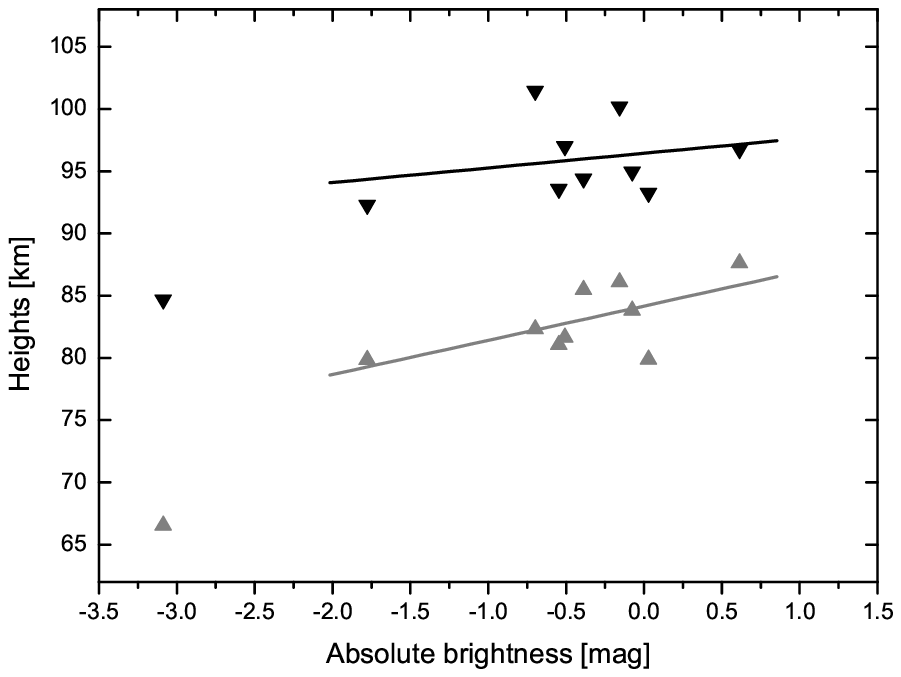}} \vspace{0.5cm}
\centerline{\includegraphics[width=8.5cm]{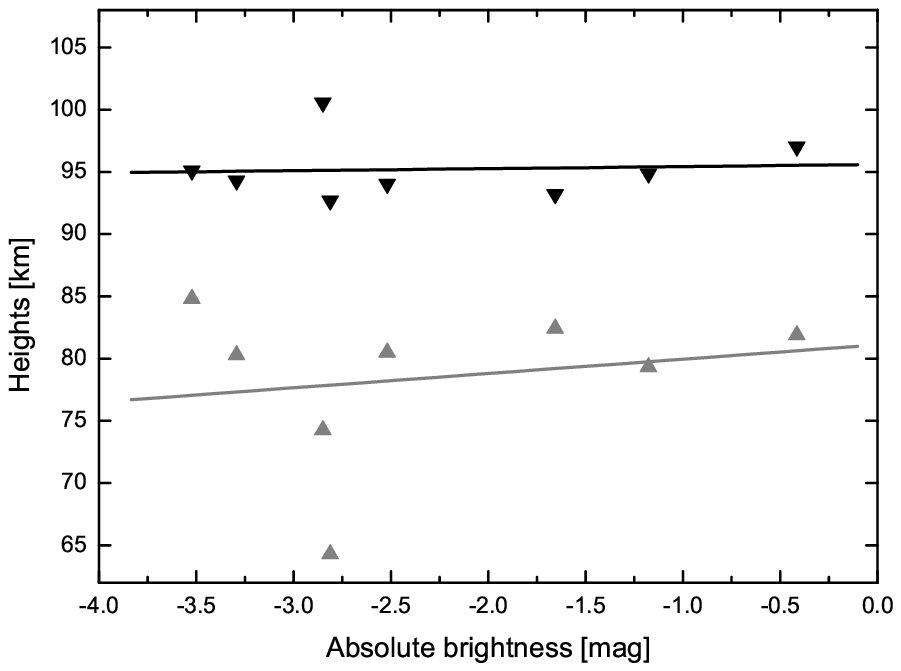}} \caption{The beginning (black
triangles) and terminal heights (gray triangles) of Geminids (upper image) and
Quadrantids (lower image) as a function of the absolute brightness.} \label{Figure
4}
\end{figure}

\section{Conclusion}

We present 10 Geminid and 8 Quadrantid heliocentric orbits of
meteors obtained by multi-station video observations done by the
Slovak Video Meteor Network and Central European Meteor Network.
The detection, data analysis and orbit derivation were made by
using the SonotaCo UFO software package. The meteor shower orbits
and their comparison with the SonotaCo database proposed parent
bodies indicate that the video observation is able to provide
relatively high quality data. Video observations offer a detection
of fainter meteors and their higher numbers in comparison with
classical photographic methods.

The coordinated video observation of meteors with amateur astronomers brings a
significant number of high quality heliocentric orbits. This cooperation has proved
to be useful due to uncertain weather in the Central Europe. In addition to current
two professional stations, future observations with amateurs might bring more
results, especially during active meteor showers or other observing campaigns.

\paragraph{Acknowledgment}
This work was supported by VEGA grant No. 1/0626/09. We greatly
appreciate the video observations and data analysis done by the
amateur astronomers of the CEMeNt network and their contribution
to this work.

{}

\end{WGNpaper}

\begin{thebibliography}{}

\bibitem{1}
\RefAut{Borovi\v{c}ka, J., Koten, P., Spurn\'{y}, P., \v{C}apek,
D., Shrben\'{y}, L., \v{S}tork, R.} \RefYear{2010}
\RefTit{Material properties of transition objects 3200 Phaethon
and 2003 EH1} \RefJnl{Icy Bodies of the Solar System, Proceedings
of the IAU Symp.} \RefVol{263}, \RefPp{218--222}

\bibitem{2}
\RefAut{Whipple, F. L.} \RefYear{1983} \RefTit{1983 TB and the Geminid Meteors}
\RefJnl{IAU Circ.} \RefVol{3881}

\bibitem{3}
\RefAut{Jenniskens P.} \RefYear{2006} \RefTit{Meteor Showers and
Their Parent Comets} Cambridge University Press, Cambridge, UK

\bibitem{4}
\RefAut{Jenniskens P.} \RefYear{2004} \RefTit{2003 EH1 Is the Quadrantid Shower
Parent Comet.} \RefJnl{Astron. J.} \RefVol{127:5}, \RefPp{3018--3022}

\bibitem{5}
\RefAut{Jacchia L.G., Whipple F. L.} \RefYear{1961} \RefTit{Precision orbits of 413
photographic meteors.} \RefJnl{Smithson. Contrib. Astrophys.} \RefVol{2},
\RefPp{181--187}

\bibitem{6}
\RefAut{Brown P., Weryk R.J., Wong D.K., Jones J.} \RefYear{2008} \RefTit{A
meteoroid stream survey using the Canadian Meteor Orbit Radar. I. Methodology and
radiant catalogue.} \RefJnl{Icarus} \RefVol{195}, \RefPp{317--339}

\bibitem{7}
\RefAut{SonotaCo} \RefYear{2009} \RefTit{Ongoing meteor work. A meteor shower
catalog based on video observations in 2007-2008.} \RefJnl{WGN, Journal of the
International Meteor Organization} \RefVol{37}, \RefPp{55--62}

\bibitem{8}
\RefAut{Southworth R.R., Hawkins G.S.} \RefYear{1963} \RefTit{Statistics of meteor
streams.} \RefJnl{Smithson. Contrib. Astrophys.} \RefVol{7}, \RefPp{261--286}

\bibitem{9}
\RefAut{Vere\v{s} P., T\'{o}th J.} \RefYear{2010} \RefTit{Analysis of the SonotaCo
video meteoroid orbits.} \RefJnl{WGN, Journal of the International Meteor
Organization} \RefVol{38:2}, \RefPp{54--57}

\bibitem{10}
\RefAut{Koten, P., Borovi\v{c}ka, J., Spurn\'{y}, P., Betlem, H., Evans, S.}
\RefYear{2004} \RefTit{Atmospheric trajectories and light curves of shower meteors}
\RefJnl{Astron. Astrophys.} \RefVol{428}, \RefPp{683--690}

\end{thebibliography}
\end{document}